# Optimized near-field optical response via adaptive tip illumination

Tao Chen[1,2], Wei Wang[1,2*], Ziyang Gan[1], Daniel Repp[3], Jinxin Zhan[4], Antony George[1], Henrik Schneidewind[2], Ulf Peschel[3], Andrey Turchanin[1], and Volker Deckert[1,2*]

[1]*Institute of Physical Chemistry and Abbe Center of Photonics, Friedrich Schiller University Jena, Helmholtzweg 4, 07743 Jena, Germany*

[2]*Leibniz Institute of Photonic Technology (IPHT), Albert-Einstein-Str. 9, 07745 Jena, Germany*

[3]*Institute of Solid State Theory and Optics, Abbe Center of Photonics, Friedrich Schiller University Jena, Max-Wien-Platz 1, 07743 Jena, Germany*

[4]*Russell Centre for Advanced Lightwave Science, Shanghai Institute of Optics and Fine Mechanics and Hangzhou Institute of Optics and Fine Mechanics, Hangzhou, China*

*Corresponding authors:
Wei Wang (w.wang@uni-jena.de)
Volker Deckert (volker.deckert@uni-jena.de).

WW and TC contributed equally.

**Abstract**

The performance of tip-enhanced optical microscopy is often limited by inefficient coupling of the excitation field to the plasmonic tip apex, as well as by thermal drift and optical aberrations. Here, we demonstrate that adaptive wavefront shaping based on Zernike mode provides a practical approach to achieving robust near-field optimisation at the tip apex. Using a sequential feedback algorithm, initially using the near-field signal, we narrow the illumination point-spread function and suppress sidelobes. This demonstrates that Zernike-mode control can be used for both aberration correction and field engineering. In tip-enhanced Raman measurements of a Janus MoSSe monolayer, conventional near-field optimisation increases the signal intensity by ~1.4-fold. A second optimisation step based directly on the Raman-band intensity yields a further 5–15-fold enhancement, depending on the specific tips used. These results establish a systematic, optics-based strategy for optimising tip fields, providing a transferable framework for improving tip-enhanced and related near-field spectroscopies.

## Introduction

Tip-enhanced Raman spectroscopy (TERS) is a form of near-field vibrational spectroscopy which localises and amplifies Raman scattering at the tip of a metallic plasmonic probe.[1-3] This enables chemical analysis with nanometre-scale spatial, and under favourable conditions, even sub-nanometre localisation.[4,5] These capabilities have made TERS an important tool for studying materials surfaces, like heterogeneous catalysis and biomaterials.[6-8] However, its broader implementation is limited by variability between tips, the need for precise optical alignment, and the sensitivity of the tips towards high-numerical-aperture excitation with respect to drift and aberrations.[9] As the local field at the tip apex is highly dependent on the geometry of the tip and how the incident optical mode couples into the plasmonic tip, even minor adjustments to the alignment can significantly impact the enhancement and compromise reproducibility. Robust control of the excitation field at the tip apex is therefore essential for reliable TERS experiments.[10]

Several strategies have been developed to improve this control. For example, Ag-tip photoluminescence has been used to facilitate hotspot localisation,[11] and illumination based on an active stabilization with a quadrant photo diode was shown.[12] Radially polarised excitation and tailored illumination geometries have been shown to improve coupling to the longitudinal field component in TERS configurations.[13,14] Correlative photo-induced force microscopy and TERS have also enabled the rapid assessment of the tip–focus interaction.[15] More recently, adaptive tip-enhanced nano-spectroscopy has used sequential wavefront shaping to optimise the nano-optical vector field at the tip apex.[16] However, the resulting phase masks segment numbers were small (12x12, each consisting of 50 x 50 pixel ), due to the intrinsically slower optimization procedure, and could not be easily interpreted in terms of physically meaningful optical modes, which limited the systematic link between wavefront correction, focal-field structure, and near-field enhancement.

In this work, we implement adaptive optics for s-SNOM and TERS based on Zernike modes projected on 800 by 800 pixel of a spatial light modulator preserving full-aperture phase control. This physically interpretable basis allows the focal field at the tip apex to be optimised fast while linking performance gains to specific modal corrections. We demonstrate that this approach improves near-field coupling and engineers the focal field by simultaneously narrowing the illumination point-spread function and suppressing sidelobes at the tip position. When combined with a second optimisation step based directly on the Raman response, this workflow provides a systematic, optics-based strategy for achieving robust and reproducible control of the tip field in TERS and related near-field spectroscopies.

## Results

### Zernike mode based adaptive wavefront shaping

As shown in Fig 1, a 532-nm continuous-wave laser was converted into a radially polarised beam (RPB) using a radial polarisation converter. This beam was then phase-modulated using a spatial light modulator before being focused onto the sample using a 1.3-NA oil-immersion objective. The reflected signal was detected by a photodiode for s-SNOM or a spectrometer/CCD for TERS. During adaptive optimization (AO), the Ag tip was fixed at the centre of the laser focus, with the demodulated $3f_0$ s-SNOM signal used as the feedback metric for sensorless wavefront control.

Starting with manual alignment (AO OFF), the coefficients of 11 Zernike modes imposed on the incident beam were used for adaption. Two sequential optimisation cycles turned out to be sufficient to reach a stable maximum in the $3f_0$ signal, after which only minor changes were observed. This procedure reproducibly increased the near-field signal across different tips, indicating improved coupling of the longitudinal focal field to the tip apex.

The optimised phase profile generated a structured pupil-field distribution rather than a simple aberration-correction pattern. This indicates that Zernike-mode control reshaped the focal field in addition to compensating for residual aberrations. This interpretation is supported by the increased tolerance of the optimised state to induced bias in selected Zernike modes, which is consistent with improved robustness of the focal coupling against alignment drift and optical aberrations. A detailed description of the AO-s-SNOM/TERS system can be found in Fig. S1 and the associated content in the Supplementary Information.

According to Knoll and Keilmann[17], the s-SNOM/TERS probe and sample EM enhancement mechanism can be understood as a dipole-dipole interaction model . Due to the strong nonlinear dependence of the scattered near-field signal on the probe-sample distance in comparison to the far-field signal, demodulating at higher harmonics improves the near-field to background ratio and leads to an improved contrast[17,18], which explains the working principle of the homodyne detection based s-SNOM used in this work. Under p-polarized illumination ($E_z$ or longitudinal component of the EM field), the EM enhancement in Eq. (1) at the tip apex is maximised[19]. The combination of the field enhancement and the Raman re-radiation induced by the TERS tip is commonly referred as the Raman enhancement factor (EF, see Eq. (2))[20]. With a highly focused RPB, $E_z$ (orthogonal to the sample surface) is the dominant field component in the laser focus[21,22]. Consequently, a RPB is at least an ideal starting point for s-SNOM/TERS experiments[21] and a higher $3f_0$-s-SNOM signal indicates a higher enhancement factor $f_e$ at the tip. Thus, the AO-s-SNOM module was used to optimize the focus starting from a standard manual aligned s-SMOM setup using the $3f_0$-s-SNOM signal (see Fig. 1b). Using tips from different batches we observed s-SNOM enhancement factors ranging between 60 to 85 at glass/air interfaces (see Supporting Information Fig. S3) and 60 to 78 at gold/air interfaces (see Supporting Information Fig. S4) indicating already a considerable improvement for the s-SNOM signal.

In the manually optimized system (noted as "AO OFF" state) the SLM was used like a flat mirror with a blazed grating phase pattern[23]. Only for the AO optimization, the SLM pattern was adapted by optimizing the coefficients of the 11 Zernike modes representing the phase profile imposed onto the initial beam (see Supporting Information Fig. S1 and Table S1). The Zernike modes are an orthogonal set of polynomials defined on the unit circle [24,25], which are commonly used to quantify phase front aberrations (such as defocus, astigmatism, spherical and higher aberration). In the current setup 800 x 800 pixel elements of the SLM were used for pattern generation of the respective Zernike modes, without any further segmentation in contrast to Ref [16]. The SLM refresh rate was < 40 ms, which represents the time limiting step for generating new Zernike mode patterns for the current system. After having optimized all coefficients, the whole cycle was repeated, and a further increase of the signal was observed. After AO optimization, the system is denoted as "AO ON".

Fig. 1c shows the phase pattern generated by the SLM after AO optimization for one TERS tip by the linear combination of the Zernike modes. The resulting intensity distribution at the objective's back-pupil plane is shown in Fig. 1d. It resembles a structured radial beam with six visible concentric rings with increasing intensity towards the outer ring. Such a complex and delicate intensity profile was unexpected, as the Zernike modes were supposed to merely correct the wavefront aberrations of the system. In order to explain this multi-ring back pupil plane pattern, we referred to the concept of Diffractive Optical Elements (DOEs)[23]. DOEs are optical devices that utilize diffraction to manipulate light, allowing for functions like beam shaping, splitting, and focusing. For example, the $Z_2^0$ Zernike mode (defocus) adapted in this work was an analogue to a classic DOE, known as a Fresnel Zone Plate (FZP)[23,26,27]. The 2nd spherical aberration term's ($Z_6^0$) profile resembled a diffractive axion. The combination of a FZP and a diffractive axicon can form ring patterns (Bessel beams) with controlled dimension[23,27]. Thus, the combined phase profile of $Z_2^0$ and $Z_6^0$ can be considered as a Fresnel axicon[23,27] that can qualitatively explain the main visible feature of the high intensity outer annulus in Fig 1 d. The simulated evolution of the optical intensity profile along the optical axis (z-axis) caused by the phase profile induced by the SLM is shown in the bottom of Fig. 1b. Finally, a concentric ring pattern with gradually reduced radius and an increasingly clear shape evolved along the z-axis. It is noteworthy that the beam intensity changes on the way from the SLM towards the objective due to diffraction (see Fig. 1b). The fitted intensity in Fig. 1e distribution resembles an inverted Laguerre–Gaussian (LG) (5,1) mode [28]. This reveals that the combined Zernike modes indeed operated more like a DOE for the generation of high order LG beams, instead of performing aberration correction in the common sense[24,25]. The application of Zernike modes in DOEs' design and analysis is well known[29]. In the present experiments, the use of Zernike modes for the direct generation of complex function DOEs was demonstrated and serves a tool to experimentally characterize such modes using tip-enhanced near-field optical methods.

Following a sequential iteration algorithm [30], the amplitudes of all Zernike modes were varied in 31 steps and the respective optimum polynomial amplitude was eventually used to obtain an optimized $3f_0$-s-SNOM signal (see also Supplementary Information Fig. S1). Fig. 2a shows a substantial increase of the $3f_0$-s-SNOM signal (> 60 after two optimization cycles), followed by a minor stabilization in the 3rd cycle. Overall, this resulted in an optimization of the s-SNOM response in less than 1 minute. In addition, the experimental results allow a comprehensive insight into the structured illumination. The respective parameters of each Zernike mode after each cycle are given in Fig. 2b. Optical aberrations can significantly affect the quality of the near-field in high-NA conditions[31,32], thereby reducing the coupling between the tip and the EM field. The impact of various optical aberrations for AO OFF in Fig. 2c and AO ON in Fig. 2d was investigated by examination of the specific phase profiles, expressed as bias terms (parameters) of the respective Zernike modes. For non-zero bias values, these modes induce aberrations as defocus ($Z_2^0$), first-order spherical aberration ($Z_4^0$), astigmatism ($Z_2^2$), coma ($Z_3^{-1}$), and tilt ($Z_1^{-1}$). As expected, inducing aberrations via respective bias changes finally led to a decay in the $3f_0$-s-SNOM signal (see Figs. 2c/d). In case of a manually optimized setup (AO OFF, Fig. 2c), the aberration induced changes were much stronger. In contrast for AO ON (see Fig. 2d, post optimization), a significant more robust focusing behavior was observed. All parameter changes of selected Zernike modes had a small impact on the $3f_0$-s-SNOM signal and the acceptance range with respect to aberrations was notably broader. In case of AO ON, the setup responded like a perfect optical system with a considerably weaker reaction to aberrations, which guarantees robust and stable alignment. This effect has already been known from adaptive two-photon microscopy[33].

As the chosen system used a 1.3 N.A. objective, it was strongly non-paraxial, and consequently the different Zernike polynomials were not fully independent (orthogonal with respect to each other). This was verified by the observation of mode crosstalk during the optimization of radial Zernike modes ($Z_6^0$, $Z_4^0$, $Z_8^0$, $Z_{10}^0$, and $Z_2^0$). Consequently, at least two cycles are required to fully optimize the system. In the experiments, the defocus parameters shifted when the Zernike mode for tilt was changed. Similar axial shifts were observed when radial Zernike modes changed. This results is consistent with observations from previous studies[31,32].

For s-SNOM/TERS, the optimization of the localized electric field at the tip apex is important. Our AO technique facilitates stable enhancement of the tip-enhanced field and is quantified by the $3f_0$-s-SNOM signal. This system can also dynamically counteract tip-field drifts, ensuring consistent TERS performance over extended durations. An optimization step can for instance activated whenever the $3f_0$-s-SNOM signal drops below 90% of the maximum value and promptly restores previous performance levels (see Fig. 2e). The deactivation (AO OFF) leads to a rapid decline. This test illustrates the sensitivity of the system to external perturbations such as thermal drift and opto-mechanical instabilities. This is particularly important in high-NA microscopy, where refractive index mismatch and inherent mechanical drift pose significant challenges to the quality of the s-SNOM and TERS signal.

**Focus Performance**

The tightly focused radially polarized beam is characterized by a dominant longitudinal field ($E_z$) and a doughnut-shaped transverse field ($E_{x/y}$) at the focal plane. The $3f_0$-s-SNOM images obtained from the RPB foci mapping of the longitudinal intensity $I_z$ show a clear maximum for both[17,34], the manual (AO OFF) and optimized (AO ON) setup (see Fig. 3a). An Airy function like behavior with concentric side-lobes was observed in the AO OFF mode, which is in good agreement with theoretical expectations for an initially homogenously illuminated objective[19,35]. Those side lobes arise from the sharp cut-off induced by the limited aperture of any diffraction-limited optical system. The resulting FWHM of the center spot was 185 nm (~ 0.45 $\lambda_0$/NA) in the AO OFF case. In the AO *ON* mode, the FWHM of the center spot was reduced to 173 nm (~ 0.43 $\lambda_0$/NA). In other words, the optimization resulted in an increase of the spatial frequency bandwidth compared to the pre-optimization reference, thus extending the spatial frequency to 1.7 times the objective's NA-defined cut-off frequency (2NA/$\lambda_0$) of 4.9 $\mu m^{-1}$ (see Fig. 3a). In the AO ON state, a remarkable 20 dB suppression of the side-lobes occurred, compared with only 2 dB for AO OFF (see inset in Fig. 3b). We attributed the suppression of the side-lobes of the AO ON state to Bessel-droplet focusing [33,36,37]. This can be achieved by an interference between multiple diffracting beams; each being generated by a bright ring in the back-pupil plane of the objective (see Fig .1b). The central parts of these beams interfere constructively, but their oscillating wings cancel each other by destructive interference resulting in a clean focal spot. In addition, the field may locally oscillate at a pace exceeding the highest Fourier component where the Fourier components outside of this segment will counterbalance these high-frequency oscillations, as the law of spatial frequency conservation must be satisfied[38]. Such fast oscillation terms are evident in the logarithmic plot of the intensity plot of AO ON (see Fig. 3b). Compared with super oscillatory focusing [39], the most significant advantage of Bessel-droplet focussing is that a sub-diffraction-limited focal spot is obtained in combination with side-lobe suppression[40-42]. According to previous studies [43-45], significant reduction of sidebands is achieved by the implementation of at least two concentric annuli (Bessel beams) in the illumination path. This multiple beam interference also causes alternating regions of low and high intensity along the optical axis, resembling "droplets" of excitation. Such subsidiary foci along the light propagation direction (z-direction) are also characteristic for a Bessel-droplet focus [40-42].

**TERS Signal Enhancement using Raman bands as AO feedback**

Finally, we demonstrate how the adaptive modal optimization can improve the TERS signal of a sample. In the experiment, a 2D Janus MoSSe (in fact: Se-Mo-S ) monolayer[46] was used. In Fig. 4a, the tip-sample-substrate model is illustrated to explain the generation of a TERS signal and its composition into a far-field and a near-field Raman (TERS) fraction. For an isolated particle with excitation at the plasmon resonance frequency, Raman EF's are modest at $EF \sim 10^5$ [8]. We modeled the EF of an Ag cone-like nanostructure with a 20° cone angle and a radius of 5 nm at the apex on a glass substrate. In the model, the distance between the tip apex and the substrate was fixed at 0.5 nm. The sample plane was assumed at 0.1 nm above the glass substrate. Interestingly, the model located the field enhancement primarily at tip apex region for AO ON, but at the tip shaft for AO OFF. The difference can be attributed to the different localized surface plasmons (LSP) eigenmodes of the tip[47]. For AO ON, only the fundamental mode is excited. The difference can be attributed to the different localized surface plasmons (LSP) eigenmodes of the tip[47]. Such a mode only exists at the outer apex area of the tip, is highly confined and leads to the excitation of strong near-fields[47]. In contrast, for AO OFF, high order LSP modes are also exited, which are bound to the shaft area and do not contribute to the effective field enhancement. According to the FDTD results, the $EF$ value at sample plane is $3.3 \times 10^5$ for AO ON, compared to the $1.2 \times 10^5$ value for AO OFF, resulting in a TERS enhancement of ~5 (approximating a $E^4$ dependency of the TERS signal).

Fig. 4b shows experimental results for the enhancement of TERS spectra using a two-step optimization and illustrates the progressive improvement of the TERS signal intensity. For a fast s-SNOM optimization as a starting point, the AO process using the $3f_0$ signal as feedback as introduced before, was applied. Subsequently, the TERS intensity of the MoSSe $A_1^1$ peak was used as feedback signal (noted as AO-$A_1^1$). Fig. 4b shows a significant intensity increase of the characteristic Stokes and anti-Stokes $A_1^1$, and the Stokes $E^2$ signals. For AO OFF, the $A_1^1$ peak was detected, while the $E^2$ was barely visible. For the first optimization step AO ON (s-SNOM only), a modest increase (1.4 factor) of the $A_1^1$ peak was observed. The second AO-$A_1^1$ TERS based optimization significantly enhanced the $A_1^1$ band. For the specific TERS tip an intensity increase by a factor of ~17 was achieved. This indicates that efficient TERS optimization eventually requires a Raman signal as feedback parameter. This effect was repeatedly observed with different tips (see Supplementary Information Fig. S7). The evolution of TERS spectra during AO-$A_1^1$ process is shown in Fig. 4.

In line with the two-step routine, the optimized Zernike mode coefficients in Fig. 4c clearly differ from the first-step $3f_0$ signal (i.e. s-SNOM) optimization. Fig. 4d finally demonstrates how Raman signals intensify stepwise enhanced through each Zernike mode optimization during AO-$A_1^1$. After normalizing intensities to their values of AO ON, an increase for all bands (except the silicon from the tip) was evident. The $A_1^1$ and $E^2$ bands increased by more than a factor of 15 and 4, respectively, while the silicon signal from probe did not directly correlate with our optimization process and even decreased. A notable rise in the $A_1^1/E^2$ ratio points to an enhanced $E_z$ field at the tip-sample region, particularly during the tuning of astigmatic modes (mode $Z_2^2$, $Z_2^{-2}$). These enhancements, evident in optimization by astigmatic modes, highlight the critical role of field refinement in Raman investigations of longitudinal modes. Our results indicate that a single $A_1^1$ feedback optimization cycle is sufficient to boost the TERS signal. This was confirmed by repeated optimization procedures using one TERS tip (see Supplementary Information Fig. S5). The AO-$A_1^1$ process effectively aligns the optical responses of various tips with the vibrational modes of interest in the sample.

## Conclusion

In summary, it is demonstrated that adaptive wavefront shaping in a Zernike-mode basis provides a fast, practical and physically interpretable route to systematically optimize the optical field at the tip apex in s-SNOM and TERS. The method effectively reduces the number of optimized degrees to the specified number of modal coefficients, here 11 Zernike modes, while preserving full-aperture phase control. In the context of near-field optical measurements, this modal approach has been shown to enhance coupling to the tip apex, suppress sidelobes, narrow the illumination point-spread function, and increase robustness against drift and optical aberrations. Consequently, it stabilises the near-field response of a given tip–sample system. In comparison to less interpretable adaptive phase masks, the Zernike basis directly links the optimisation to physically meaningful modal corrections and enables a more efficient and hands-free optimisation workflow for high-NA near-field experiments. When combined with a second optimisation step based directly on selected TERS-signal intensities, this strategy has been shown to further improve TERS performance beyond that achieved by near-field optimisation alone. The results obtained establish a systematic optics-based framework for faster, robust and reproducible tip-field control in s-SNOM, TERS and related tip-enhanced spectroscopies.

## Acknowledgements

The authors acknowledge gratefully funding via different agencies: TC, UP, AT and VD are supported by the Deutsche Forschungsgemeinschaft (DFG) SFB 1375 NOA (No. 398816777). WW, AT and VD were supported by the DFG, TRR 234 "CataLight" Projects C4 and Z2 (No. 364549901). WW received additional support by the IPHT Innovation Project 2021/2022 (690082). All simulations were performed on the HPC-cluster supported by EFRE Programm and Freistaat Thüringen, projects EU-0V/2020-59 (2019 FGI0017) and EU-0V/2023-1 (2022 FGI 0004). WW thanks Dr. Oleg A. Egorov for the support with the simulation. WW thanks Dr. David Albertini and Dr. Adrianos Sidiras Galante for the open source Zi[2] software (https://github.com/UnlikelyBuddy1/ZI2) for the data acquisition of s-SNOM signals. This work is supported by the BMBF, funding program Photonics Research Germany („SARS-CoV-2Dx", FKZ: 13N15745) and integrated into the Leibniz Center for Photonics in Infection Research (LPI). The LPI initiated by Leibniz-IPHT, Leibniz-HKI, UKJ and FSU Jena is part of the BMBF national roadmap for research infrastructures.

## Author Contributions Statements

V.D. conceived the idea, and organized, coordinated, and supervised the project. T.C. performed all measurements and collected the complete experimental dataset (excluding simulation data). W.W and T.C. performed and optimized the s-SNOM setup. J.Z. contributed to discussions to the s-SNOM setup. U.P., D.R., and W.W. performed the FDTD simulations. A.T., A.G., and Z.G. performed material growth and structural characterization. H.S. performed TERS tip preparation. V.D., U.P., T.C., and W.W. interpreted the results and proposed the mechanism. T.C. and W.W. wrote the manuscript. V.D. supervised and reviewed the manuscript. V.D., U.P., A.T., T.C., J.Z. and W.W. discussed and revised the manuscript.

## Competing Interests Statement

The authors declare no competing financial interest.

## Figures & Captions

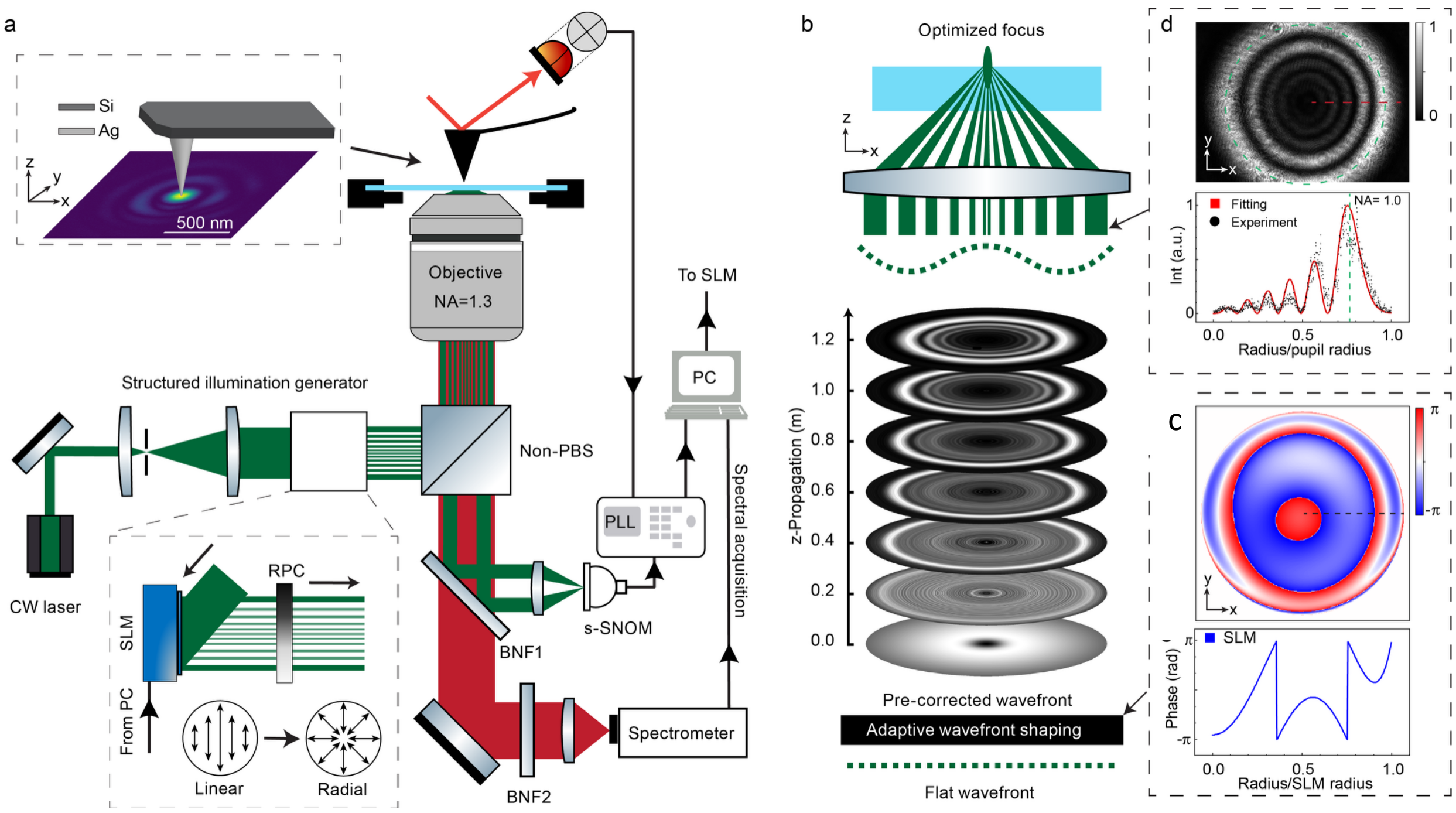

**Fig. 1: Adaptive-optimization scheme for s-SNOM and TERS.**

**a**, Schematic of the adaptive-optimization s-SNOM/TERS setup. Wavefront shaping is optimized using either the s-SNOM signal or the intensity of a selected Raman band as the feedback metric. **b,** Schematic of the optimized focal field generated by multi-ring radially polarized illumination, and simulated beam profile along the optical axis. **c,** Experimentally optimized phase distribution at the spatial light modulator plane in the AO ON state, shown as a phase map and radial profile. **d,** Experimentally measured field amplitude at the objective back focal plane in the AO ON state, shown as an amplitude map and radial profile; dots indicate experimental data and the red line indicates the fit. The outermost ring overlaps with NA = 1.0 (green dashed line). CW, continuous wave; SLM, spatial light modulator; RPC, radial polarization converter; non-PBS, non-polarizing beam splitter; BNF, Bragg notch filter; PLL, phase-locked loop

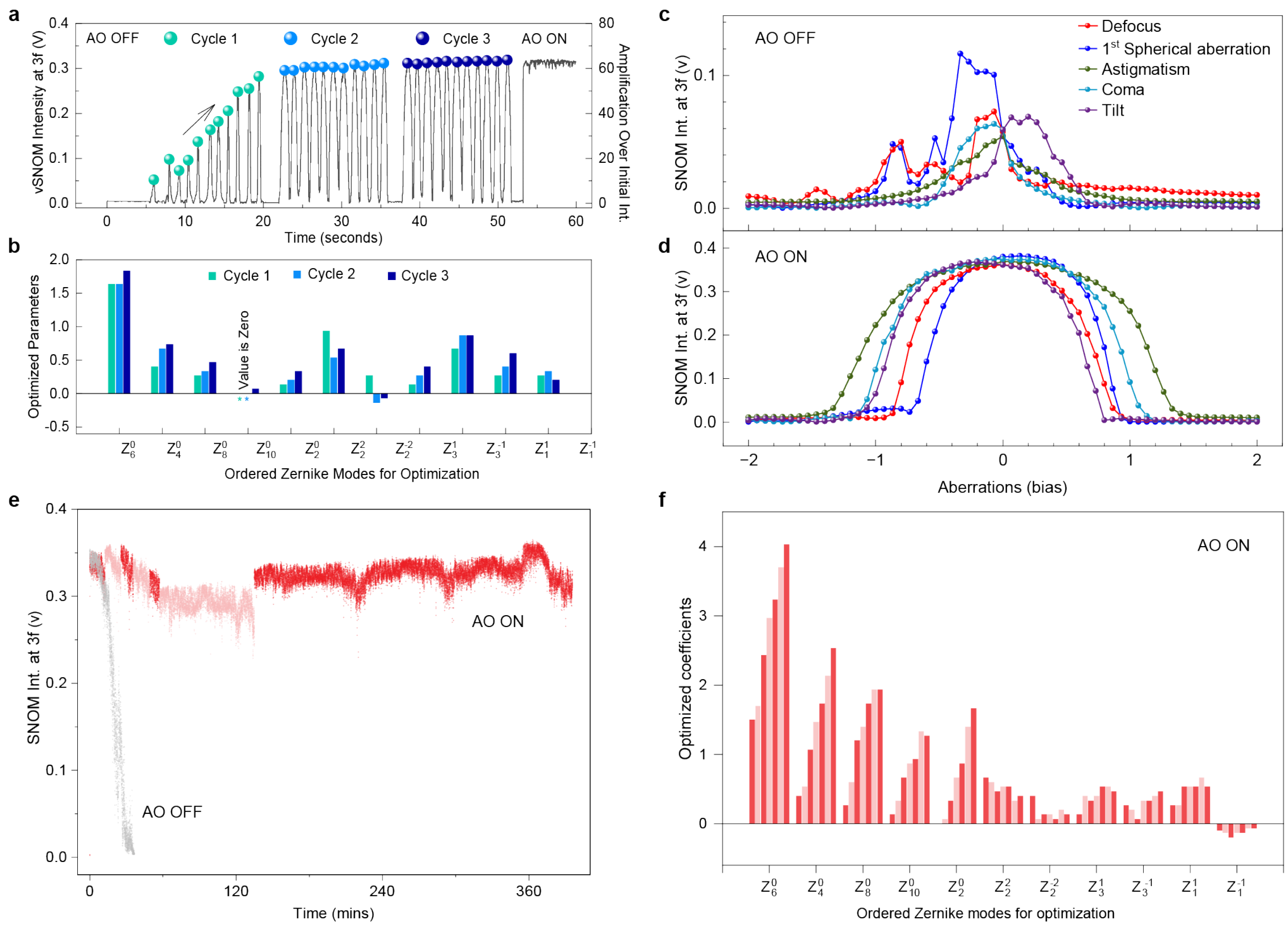

**Fig. 2: Adaptive optimization and stabilization of the longitudinal near field.**

**a,** Sequential increase of the $3f_0$ s-SNOM signal over three adaptive-optimization cycles. **b,** Optimized Zernike-mode coefficients after each cycle. **c,d,** Response of the $3f_0$ s-SNOM signal to induced aberrations in the AO OFF (**c**) and AO ON (**d**) states. The manually aligned system (AO OFF) and the adaptively optimized system (AO ON) were each perturbed by applying defined phase biases with the spatial light modulator, including defocus, first-order spherical aberration, astigmatism, coma and tilt. **e,** Stabilization of the $3f_0$ s-SNOM signal during repeated adaptive re-optimization. Red and light-red segments indicate periods after reactivation of the adaptive correction when the signal had fallen below 90% of its maximum value; grey indicates the corresponding signal evolution without adaptive correction. **f,** Zernike-mode coefficients during the re-optimization periods shown in **e.** For clarity, the intermediate optimization steps are omitted in **e.**

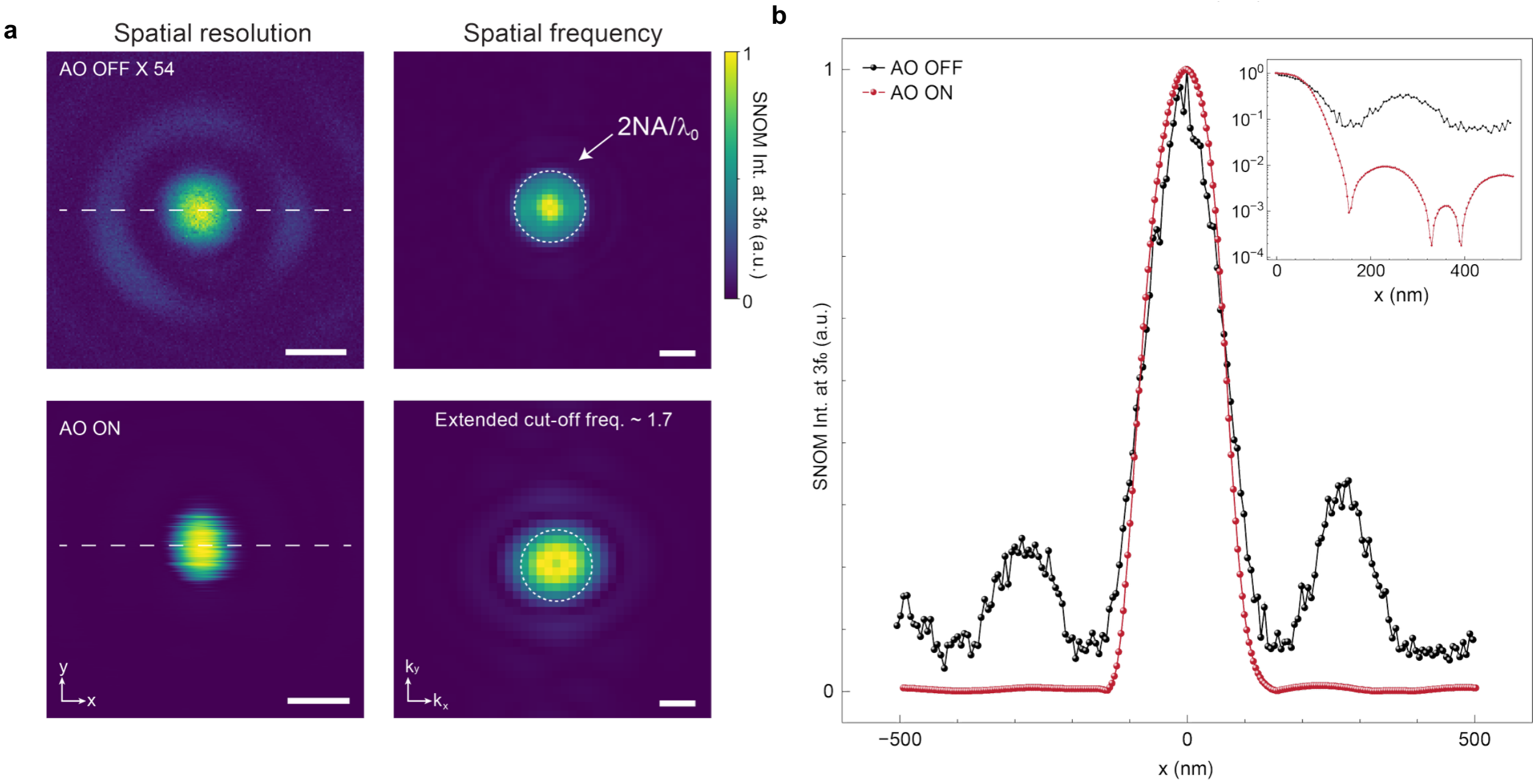

**Fig. 3: Real-space and k-space mapping of the longitudinal near field before and after adaptive optimization.**

**a,** s-SNOM maps of the focal-plane near-field signal in real space and the corresponding k-space distributions in the AO OFF and AO ON states. Dashed circles indicate the cut-off frequency $(2\mathrm{NA}/\lambda_0)$. Scale bars, 200 nm (real space) and 5 μm$^{-1}$ (k-space). **b,** Line profiles of the focal-plane s-SNOM signal in the $xy$ plane for the AO OFF and AO ON states. Inset, logarithmic representation of the same profiles highlighting the sidelobes.

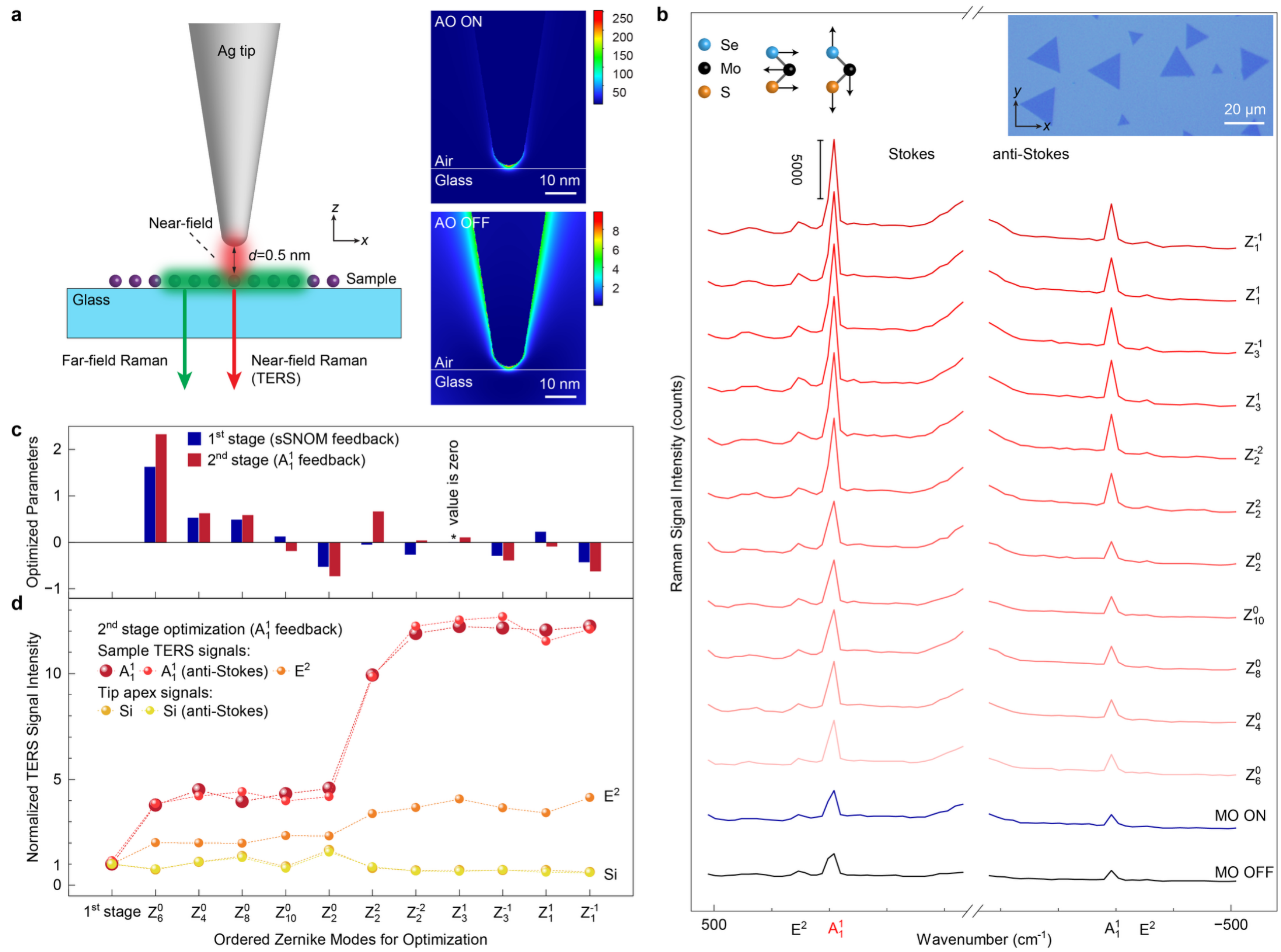

**Fig. 4: TERS enhancement of a MoSSe monolayer using a two-step adaptive optimization.**

**a,** Schematic of the epi-illumination TERS geometry. The Ag tip is modelled as a cone with a 20° opening angle and a 5 nm apex radius, positioned 0.5 nm above the glass interface. The Raman signal comprises a far-field contribution (green) and a near-field TERS contribution (red). Inset, modelled $|E/E_0|^4$ enhancement factor for the AO OFF and AO ON states. **b,** Background-corrected Stokes and anti-Stokes TERS spectra acquired ($t_{acq}$=0.5 s; $P_{@sample}$=22µW ) during the two-step optimization procedure. The first step uses the $3f_0$ s-SNOM signal as the feedback metric (AO ON), and the second step uses the intensity of the $A_1^1$ Raman mode as the feedback metric, referenced to the spectrum after s-SNOM optimization. Inset, optical image of a triangular MoSSe flake. Scale bar. **c,** Zernike-mode coefficients obtained after optimization with s-SNOM feedback and with $A_1^1$-mode Raman feedback. **d,** Evolution of the principal Raman bands during the second optimization step.

## Method

### System configuration:

As shown in Fig.1a, the system is composed of three major components: TERS, s-SNOM and AO module. A Spatial Light Modulator (SLM) is illumined by a linearly polarized laser beam at 532 nm ($\lambda_0$) and 22 µW (Cobolt 04-01 series) and the resulting phase modulated field is converted to a radially polarized beam (RPB) using a radial polarization converter. A NA=1.3 objective lens is used for focusing the incident RPB on to the sample plane/tip apex and also for collecting the reflected/scattered signal. A notch filter separates the 532 nm reflection and Rayleigh scattered light from the Raman signal. The s-SNOM fraction is guided to a photodiode detector, the Raman signal to a spectrometer. To suppress far-field background signal, the input signal from the photodiode is demodulated at the 3rd ($3f_0$) harmonic of the oscillation frequency $f_0$ of the cantilever to obtain the s-SNOM signal using a lock-in amplifier. Raman spectra were detected by a home built system. Tip-feedback, scanning was maintained by an AFM system (Nanowizard 2, Bruker-JPK). More details regarding the setup can be found in the supporting information.

### Raman enhancement factor:

For TERS/s-SNOM, the EM enhancement factor around tip apex can be expressed as:

$$\sqrt{f_e} = |\boldsymbol{E}/\boldsymbol{E}_0| \tag{1}$$

In Eq. (1), $\boldsymbol{E_0}$ denotes the incident field and $\boldsymbol{E}$ denotes electric field in the vicinity of the tip.

According to Kerker [20], the combination of field enhancement and Raman re-radiation induced by the metal tip is commonly referred as the $|\boldsymbol{E}/\boldsymbol{E_0}|^4$ approximation, where the Raman signal enhancement factor EF is defined as following:

$$EF = \frac{|\boldsymbol{E_{exc}}|^2}{|\boldsymbol{E_0}|^2} \cdot \frac{|\boldsymbol{E_{em}}|^2}{|\boldsymbol{E_0}|^2} \approx \left|\frac{\boldsymbol{E}}{\boldsymbol{E_0}}\right|^4 \tag{2}$$

In Eq. (2), the enhancement of the electric field exciting the sample ($\boldsymbol{E_{exc}}$) and Raman scattering ($\boldsymbol{E_{em}}$) are normalized to the $\boldsymbol{E_0}$.